# Probing Cluster Potentials through Gravitational Lensing of Background X-Ray Sources[*]

A. Refregier[1], and A. Loeb[2]

[1] Columbia Astrophysics Laboratory, 538 W. 120th Street, New York, NY 10027
[2] Astronomy Department, Harvard University, Cambridge, MA 02138
[*] to appear in Proc. of "Röntgenstrahlung from the Universe", Würzburg, Germany, Sept. 1995

**Abstract.** We examine the gravitational lensing effect of a foreground galaxy cluster on the number count statistics of background X-ray sources. The lensing produces a deficit in the number of resolved sources in a ring close to the critical radius of the cluster. This deficit could be detected at the $\sim 3\sigma$ level with the AXAF-ACIS camera.

## 1. Introduction

Several complementary methods are routinely used to probe the gravitational potential of clusters of galaxies. These methods are based on: (i) the dynamics of the galaxy members of the cluster, (ii) the X-ray brightness distribution of the cluster gas, and (iii) the gravitational lensing of background galaxies. While the first two methods rely on simplifying assumptions (e.g. spherical symmetry, hydrostatic equilibrium), gravitational lensing can directly measure the surface mass distribution in clusters.

Previous discussions of strong and weak gravitational lensing by clusters have focused on distortions in the optical images of background galaxies (see, e.g. Kaiser 1995, for a review). Here, we study the effect of a cluster potential on the number count statistics of background X-ray sources. Since most X-ray sources which constitute the cosmic X-Ray Background (XRB) are AGN (Fabian & Barcons 1992), they should appear as point sources and be potentially more highly magnified than the extended images of galaxies. In particular, the cluster lens can be used as a *natural telescope* to study the faint end of the $\log N$-$\log S$ relation for the sources which account for the XRB. The full details of this work can be found in Refregier & Loeb (1996). Throughout this paper we assume $\Omega = 1$ and $H_0 = 50$ km s$^{-1}$ Mpc$^{-1}$.

## 2. Models

We model the cluster mass distribution as a Singular Isothermal Sphere (e.g., Schneider et al. 1991) which successfully approximates known lenses (e.g., Grossman & Narayan 1989). Gravitational lensing causes the background sources behind the cluster to be magnified by a factor $\mu$, which depends on angular position as $\mu(\theta) = 1/(1 - |\alpha_0/\theta|)$, where $\theta$ is the angle between the cluster center and the lensed image of the source, and $\alpha_0$ is the critical angle. In general, $\alpha_0$ depends on the redshifts of the cluster $z_{cl}$ and the source $z_s$, and on the line-of-sight velocity dispersion $\sigma_v$ of the cluster. For $\theta \gg \alpha_0$, $\mu$ is close to unity and the sources are only weakly affected by lensing. However, for $\theta \sim \alpha_0$, the magnification is large.

For a nearby cluster ($z_{cl} \lesssim 0.2$) and for sources with $z_s \gtrsim 0.5$, $\alpha_0$ is only a weak function of $z_s$. To first approximation, we therefore model the background X-ray sources as a set of randomly distributed points with $z_s = \infty$. The flux distribution of the sources is determined by their $\log N$–$\log S$ relation which we model using three broken power laws subject to the following observational constraints (Hasinger et al. 1993): (i) the model provides the observed ROSAT counts for $S(0.5\text{-}2\text{keV}) < 2 \times 10^{-15}$ ergs s$^{-1}$ cm$^{-2}$; (ii) the model lies within the limits from the ROSAT fluctuation analysis; and (iii) the integrated intensity from the sources produces 100% of the extragalactic XRB in the ROSAT band. Figure 1 shows the observed ROSAT $\log N$–$\log S$ relation and the fluctuation analysis limits. Among the many possible extrapolations, we consider here a case with a moderate slope. This model is shown as the solid line in Figure 1.

## 3. Effect of lensing on background sources

Cluster lensing affects background sources in two ways: their apparent fluxes $S$ are magnified by a factor $\mu$, and their surface number density $N$ is diluted by the same factor. As a result, the observed number of sources per solid angle per unit flux at a given apparent flux $S$ scales as $\frac{dN}{dS}|_S = \frac{1}{\mu^2}(\frac{dN}{dS})'|_{S/\mu}$, where unprimed (primed) quantities refer to the situation with (without) lensing. Figure 1 shows the observed $\log N$–$\log S$ relations obtained by integrating the differential counts, for two values of $\mu$. At fluxes fainter than $S_* \simeq 4 \times 10^{-15}$ ergs s$^{-1}$ cm$^{-2}$, the source dilution effect dominates over the magnification effect and $N(<S)$ decreases. The opposite is true for $S > S_*$. We therefore expect that in regions with high $\mu$, e.g. in a ring close to the critical radius $\alpha_0$, the number of sources $N(>S_r)$ with fluxes above a given detection threshold, $S_r$, will be reduced if $S_r < S_*$.

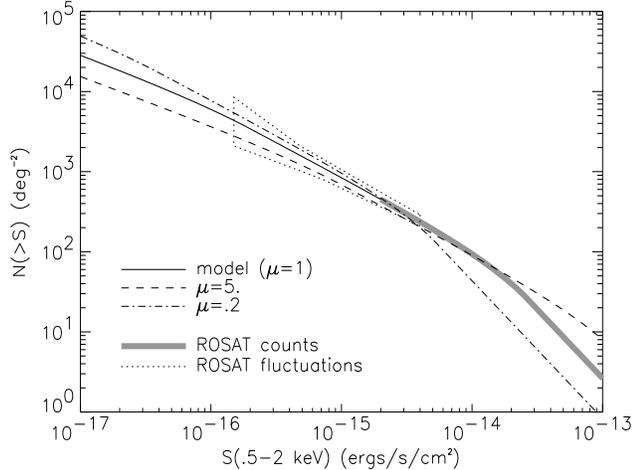 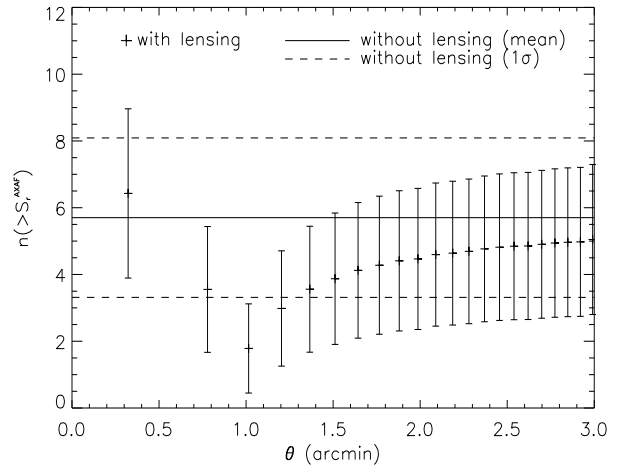

**Fig. 1.** The log $N$–log $S$ relation for background X-ray sources in the ROSAT (.5-2 keV) band. The thick grey line and the dotted line show the observed ROSAT counts and fluctuation analysis limits, respectively (Hasinger et al. 1993). The thin solid line represents our chosen extrapolation without lensing. The dashed and dot-dashed line show the observed relation when the magnification $\mu$ is 5 and 0.2, respectively.

**Fig. 2.** Number of sources which could be resolved by AXAF in concentric rings of equal area (1.31 square arcminutes), for the cluster example discussed in the text. The solid line and the crosses represent the source counts without and with lensing. The error bars and the dashed line represent $1\sigma$ Poisson standard deviations for these cases.

## 4. Observability

To see whether the lensing effect could be detected in the future, we consider an example of a cluster at $z_{cl} = 0.1$ with $\sigma_v = 1500$ km sec$^{-1}$. The resulting critical angle $\alpha_0$ is about $1'$. We assume that the X-ray emitting gas follows a spherical King profile (e.g. Sarazin, 1988) with a core radius of 0.5 Mpc, $\beta = 0.65$ (corresponding to a gas temperature of $kT_g = 11$ keV), and a total luminosity of $L_x = 1.8 \times 10^{45}$ erg s$^{-1}$ in the 0.5-4 keV band. We further assume that the cluster X-ray emission has a thermal bremsstrahlung spectrum with a temperature $T_g$. The ACIS instrument on board AXAF will achieve a FWHM angular resolution of $0.5''$ (Forman 1995). Taking the sources to have a spectral power-law index of 0.6, the AXAF-ACIS $2.5\sigma$ detection threshold for sources *behind* the cluster is $S_r^{AXAF}(0.2-8\text{keV}) \simeq 2.5 \times 10^{-17}$ ergs s$^{-1}$ cm$^{-2}$ for an exposure time of $10^6$ seconds.

Figure 2 shows the expected counts as a function of radius for the sources which can be resolved by AXAF. The sources have been binned in concentric rings of area 1.31 square arcminutes around the cluster center. The expected deficit of sources around the critical radius $\theta = \alpha_0 \simeq 1'$ is clearly noticeable and is significant at the $\sim 3\sigma$ level in this example. The cluster emission degrades the AXAF detection threshold. In its absence, the deficit would have had a significance of $\sim 3.5\sigma$.

Observation of cluster lensing of X-ray sources requires a rather long AXAF integration time. A positive detection would complement current lensing studies, especially in the strong lensing regime. The count deficit shown in Figure 2 also depends on the choice of the log $N$–log $S$ extension at faint fluxes. Lensing could therefore uncover the origin of the XRB, not only by magnifying otherwise unresolved X-ray sources, but also by constraining the faint end slope of its log $N$–log $S$ relation.

*Acknowledgements.* We thank D.J. Helfand for useful comments and for his support of AR. We are also grateful to M. Bartelmann and W. Forman for insightful discussions. This work was supported in part by NASA grant NAGW2507 (for AR) and by the NASA ATP grant NAG5-3085 (for AL). This work appears as report number 586 of the Columbia Astrophysics Laboratory.